\documentclass[]{aa}  

\usepackage{natbib}
\bibpunct{(}{)}{;}{a}{}{,}
\usepackage{graphicx}
\usepackage{revsymb}	
\usepackage{amsmath}	
\usepackage[usenames]{color}	
\usepackage{txfonts}
\usepackage{amssymb}
\usepackage{color}
\usepackage[justification=centering]{caption}
\usepackage{geometry} 
\usepackage[parfill]{parskip} 
\usepackage{amssymb}
\usepackage{slantsc}
\usepackage{array}
\usepackage{amsfonts}
\usepackage[colorlinks=true,linkcolor=black, urlcolor=black, citecolor=black]{hyperref}
\usepackage{sidecap}
\usepackage{graphicx}
\usepackage{xcolor}
\usepackage{nicefrac}
\usepackage{subfigure}
\usepackage{epsfig}
\usepackage{url}
\usepackage{breakurl}

\colorlet{rouge}{red!70!darkgray}

\begin{document}
\title{In-depth study of $16$CygB using inversion techniques}
\author{G. Buldgen\inst{1} \and S.J.A.J. Salmon\inst{1} \and D. R. Reese\inst{2}\and M. A. Dupret\inst{1}}
\institute{Institut d’Astrophysique et Géophysique de l’Université de Liège, Allée du 6 août 17, 4000 Liège, Belgium \and LESIA, Observatoire de Paris, PSL Research University, CNRS, Sorbonne Universités, UPMC Univ. Paris 06, Univ. Paris Diderot, Sorbonne Paris Cité, 5 place Jules Janssen, 92195 Meudon Cedex, France}
\date{April, 2016}
\abstract{The $16$Cyg binary system hosts the solar-like Kepler targets with the most stringent observational constraints. Indeed, we benefit from very high quality oscillation spectra, as well as spectroscopic and interferometric observations. Moreover, this system is particularly interesting since both stars are very similar in mass but the $A$ component is orbited by a red dwarf, whereas the $B$ component is orbited by a Jovian planet and thus could have formed a more complex planetary system. In our previous study, we showed that seismic inversions of integrated quantities could be used to constrain microscopic diffusion in the $A$ component. In this study, we analyse the $B$ component in the light of a more regularised inversion.}
{We wish to analyse independently the $B$ component of the $16$Cyg binary system using the inversion of an indicator dedicated to analyse core conditions, denoted $t_{u}$. Using this independent determination, we wish to analyse any differences between both stars due to the potential influence of planetary formation on stellar structure and/or their respective evolution.}
{First, we recall the observational constraints for $16$CygB and the method we used to generate reference stellar models of this star. We then describe how we improved the inversion and how this approach could be used for future targets with a sufficient number of observed frequencies. The inversion results were then used to analyse the differences between the $A$ and $B$ components.}
{The inversion of the $t_{u}$ indicator for $16$CygB shows a disagreement with models including microscopic diffusion and sharing the chemical composition previously derived for $16$CygA. We show that small changes in chemical composition are insufficient to solve the problem but that extra mixing can account for the differences seen between both stars. We use a parametric approach to analyse the impact of extra mixing in the form of turbulent diffusion on the behaviour of the $t_{u}$ values. We conclude on the necessity of further investigations using models with a physically motivated implementation of extra mixing processes including additional constraints to further improve the accuracy with which the fundamental parameters of this system are determined.}{}
\maketitle
\section{Introduction}\label{SecIntro}

In a previous paper \citep{BuldgenCyg}, we studied the binaries $16$CygA and $16$CygB using the full Kepler dataset from \citet{Davies}. The system is in fact more complex since a red dwarf orbits the $A$ component and a Jovian planet orbits the $B$ component \citep{Cochran, Holman, Hauser}. We carried out a forward modelling process of both stars without taking into account binarity as a constraint and used our inversion techniques to further constrain their fundamental parameters, and demonstrated the importance of microscopic diffusion. The inversion technique provided strong constraints on the chemical composition and mixing of $16$CygA, the brightest of the two components. However, when carrying out the same inversion for $16$CygB, we faced the problem of the amplification of the observational error bars. The problem is well-known in the context of inversions, since the results are always a trade-off between amplifying the errors and fitting the target function of the inversion \citep{Pijpers}. In the context of asteroseismology, since more weight has to be given to the fit of the target function due to the small number of observed frequencies compared to the solar case, we are always limited in terms of error amplification. Trying to reduce the error bars by amplifying the trade-off parameters can result in a significant reduction of the quality of the fit, thus implying that what is gained by reducing the propagation of observational error bars is lost due to the poor quality of the averaging kernel.

In the following sections, we re-analyse the trade-off problem of $16$CygB and show that the seismic information is sufficient to analyse this star independently with the $t_{u}$ indicator. To explain the trend seen with the inversion, we try unsuccessfully to restore the agreement by modifying the surface chemical composition of this star. Since this leads to inconsistencies with the $16$CygA results of our previous paper, we analyse the potential necessity of an additional mixing process, which has already been mentioned to explain the lithium depletion in this star \citep{Deal}. We emphasize that the solution we propose for consistency with the inversion result is hypothetical and is subject to the same limitations and model-dependencies as our previous study on $16$CygA. We compute models using a parametrised approach of the extra mixing which should not be considered as a physical solution but rather a hint that a certain amount of mixing is required in deep regions of the B component in order to reconcile the modelling of both components.

The paper is structured as follows, we start by briefly presenting additional reference models in Sect \ref{SecRefMod}. We then present our inversion results as well as the regularisation in Sect \ref{SecInvRes}. These results are further analysed and discussed in Sect \ref{SecDisc} in light of the possible necessity for extra mixing in $16$CygB. We then conclude with the implications and perspectives of this study in Sect \ref{SecConc}. 
\section{Reference models}\label{SecRefMod}
In this section, we will describe the forward modelling process that has been carried out to obtain the reference models for the inversion. The process has been already described in \citet{BuldgenCyg}, but we recall it here for the sake of clarity. Nevertheless, the number of models computed has been increased to improve the diagnostic process of the inversion and to ensure unbiased results.

In practice, we computed these models independently from the modelling of $16$CygA presented in our previous paper. We used the frequency spectrum from \citet{Davies}, which was based on $928$ days of Kepler data \footnote{The frequency tables are public and can be found at the url: \url{http://mnras.oxfordjournals.org/lookup/suppl/doi:10.1093/mnras/stu2331/-/DC1}.}. A Levenberg-Marquardt algorithm was used to determine the optimal set of free parameters for our models. We used the CLES stellar evolution code and the LOSC oscillation code \citep{ScuflaireCles, ScuflaireLosc}. The stellar models used the CEFF equation of state \citep{CEFF}, the OPAL opacities from \citet{OPAL} supplemented at low temperatures by the opacities of \citet{Ferguson}. The nuclear reaction rates are those from the NACRE project \citep{Nacre}, including the updated reaction rate for the $^{14}\mathrm{N}(p,\gamma)^{15}\mathrm{O}$ reaction from \citet{Formicola} and convection was implemented using the classical, local mixing-length theory \citep{Bohm}. We also used the implementation of microscopic diffusion from \citet{Thoul}, for which three groups of elements are considered and treated separately: hydrogen, helium and the metals (all considered to have the diffusion speed of $^{56}Fe$). No additional transport mechanism, beside microscopic diffusion, was included in the models. No surface correction of the individual frequencies was used in this study since we used quantities that are naturally less sensitive to these effects.

Moreover, since the inversion results for $16$CygA implied that microscopic diffusion had to be included in the stellar models and since both stars are very similar, we considered that we had to include atomic diffusion in the models of $16$CygB. We also emphasize that obtaining consistent results in age for both components is impossible if one considers that one component of the binary system is subject to microscopic diffusion effects while the other is not. Yet, we also want to stress that the implementation of microscopic diffusion has its own uncertainties. First, we consider here the implementation from \citet{Thoul} which considers only three components to the mixing; secondly, in their own paper, \citet{Thoul} consider the diffusion velocities obtained to be accurate within approximately $15 \%$; thirdly, it may be possible that radiative accelerations play a role in competing with gravitational settling effects. Thus, the use of microscopic diffusion as a solution to be consistent with the inversion results for $16$CygA is a first hypothesis of this study. It does not mean that another combination of mixing processes could not successfully reproduce the trends previously seen with the inversion technique for this star.

In this study, we substantially increased the number of reference models used to carry out the inversions for $16$CygB but did not use any hypothesis on the chemical composition of this star. In fact, surface chemical composition differences between the $A$ and $B$ components have been claimed by \citet{Tucci} when carrying out a differential spectroscopy analysis between both stars. Moreover, although the centroid of the present surface helium abundance, $Y_{f}$, interval found by \citet{Verma} is the same, the scatter is larger for the $B$ component, and if microscopic diffusion is included in the stellar models, it is also clear that surface chemical composition differences will be seen since this mixing will not have the same efficiency for stars of different masses\footnote{The differences due to diffusion should nonetheless remain small.}.

Nevertheless, it should be noted that chemical composition differences between $16$CygA and $16$CygB are still under some debate since their existence has been claimed by \citet{Ramirez} and \citet{Tucci} as well as by previous studies \citep[see][]{Deliyannis} but could not be confirmed by \citet{Schuler}. In \citet{Tucci}, one finds $\left[ \mathrm{Fe/H} \right]_{A}=0.101 \pm 0.008$ and $\left[ \mathrm{Fe/H} \right]_{B}=0.054 \pm 0.008$ whereas \citet{Schuler} finds $\left[ \mathrm{Fe/H} \right]_{A}=0.07 \pm 0.05$ and $\left[ \mathrm{Fe/H} \right]_{B}=0.05 \pm 0.05$. These results are not totally incompatible, and what is more striking is the difference in error bars between various studies.

Moreover, these values depend on the reference solar metallicity assumed in the study since the observational constraint provided is the $\left[ \mathrm{Fe/H} \right]$ value which must be translated in a $\frac{Z}{X}$ value using the sun as a reference. In our previous paper, we used the most recent abundance tables given by AGSS09 \citep{AGSS} and found that they led to a better agreement with the inversion results for $16$CygA. In this study, we computed most models with the AGSS09 abundances but also used some models with the older GN93 abundances \citep{GN93}. We explain our motivations for using such models in Sect. \ref{SecDisc}.
\begin{table*}[t]
\caption{Summary of observational properties of the system $16$CygA B used in this study.}
\label{tabObs}
  \centering
\begin{tabular}{r | c | c }
\hline \hline
 & \textbf{$16$CygB} & \textbf{References} \\ \hline
\textit{R ($\mathrm{R_{\odot}}$) }& $1.12 \pm 0.02$ &  \citet{White} \\ 
\textit{$T_{\mathrm{eff,spec}}$ ($\mathrm{K}$)} & $5751 \pm 6$ &\citet{Tucci} \\ 
\textit{$T_{\mathrm{eff,phot}}$ ($\mathrm{K}$)} & $5809 \pm 39$ & \citet{White} \\ 
\textit{$L$ $(\mathrm{L_{\odot}})$} & $1.27 \pm 0.04$ & \citet{Metcalfe} \\
\textit{[Fe/H] (dex)} &$0.052 \pm 0.021$ & \citet{Ramirez} \\
\textit{$Y_{f}$} &$\left[0.218, 0.260\right]$ & \citet{Verma} \\
\textit{$<\Delta \nu>$} ($\mu \mathrm{Hz}$) &$117.36 \pm 0.55$ & \citet{Davies} \\
\hline
\end{tabular}
\end{table*}
We summarise the observational constraints used for $16$CygB in table \ref{tabObs} and the fundamental parameters obtained for some of the reference models in table \ref{tabresB}. In this table, we also recall the intervals from the forward modelling process of $16$CygA obtained previously. The forward modelling war carried out starting from various initial conditions with the Levenberg-Marquardt algorithm. The set-up of the minimization process was the following:
\begin{itemize}
\item Constraints: individual small frequency separations $d_{0,2}$ $d_{1,3}$, inverted mean density $(\bar{\rho})$ for which conservative error bars of $0.005$ $g/cm^{3}$ were considered, acoustic radius $(\tau)$ for which conservative error bars of $30$ $s$ were considered, present surface metallicity $(Z_{f}/X_{f})$ from \citet{Ramirez}, present surface helium abundance $(Y_{f})$ from \citet{Verma} and the effective temperature from \citet{Tucci}, for which we considered error bars of $30$K.
\item Free parameters: Mass, age, initial hydrogen abundance $(X_{0})$, initial abundance of heavy elements $(Z_{0})$, mixing-length parameter $(\alpha_{MLT})$. 
\end{itemize}
In total, we had $5$ free parameters for $31$ constraints. In addition to these constraints, we checked the values of the luminosity $L$, surface gravity $\log g$ and radius $R$ after the forward modelling to see if they were consistent with the constraints from the literature. Models which were completely inconsistent with these additional constraints were disregarded. An additional comment should be made on some error bars used in the forward modelling. Firstly, we considered the errors from \citet{Tucci} to be unrealistic and assumed a conservative $30$K error bar which is already very accurate but more consistent with other studies. Secondly, both the inverted mean density and acoustic radius are known to have underestimated error bars with the SOLA method, from the multiple hare and hounds we performed to calibrate the inversion techniques, we noticed that a error bars of $0.5\%$ were to be expected as a conservative error bar for the inverted values of the mean density. For the acoustic radius, the precision has to be assessed from the dispersion of the inverted values, in this particular case this lead to a precision of around $0.7 \%$ was achieved. Consequently, we used these conservative error bars in the Levenberg-Marquardt algorithm rather than the error bars derived directly from the SOLA method.

\begin{table*}[t]
\caption{Parameters of the reference models of $16$CygB}
\label{tabresB}
  \centering
\begin{tabular}{r | c | c}
\hline \hline
 & \textbf{Reference $16$CygB models} &\textbf{Reference $16$CygA models} \\ \hline
 \textit{Mass ($\mathrm{M_{\odot}}$)}& $0.93$-$1.05$ & $0.96$-$1.08$\\
\textit{Radius ($\mathrm{R_{\odot}}$)}& $1.07$-$1.13$  & $1.19$-$1.24$\\ 
\textit{Age ($\mathrm{Gyr}$)} &$6.97$-$8.47$ &$6.90$-$8.30$\\ 
\textit{$L_{\odot}$ $(\mathrm{L_{\odot}})$} & $1.05$-$1.25$ & $1.48$-$1.66$\\
\textit{$Z_{0}$} & $0.0165$-$0.0194$ & $0.0155$-$0.0210$\\
\textit{$Y_{0}$} &$0.25$-$0.32$ & $0.250$-$0.299$\\
\textit{$\alpha_{\mathrm{MLT}}$} & $1.70$-$1.86$ & $1.67$-$1.97$\\
\textit{$D$} & $0.5$-$1.1$ & $0.0-1.1$\\
\hline
\end{tabular}
\end{table*}
We can see that the scatter of fundamental parameters is very similar to that obtained for $16$CygA. However, we only give the results for models including diffusion in table \ref{tabresB}, as can be seen by looking at the values of the $D$ parameter. This parameter is related to the implementation of diffusion we use, it is a multiplicative factor of the microscopic diffusion velocities such that if $D=1.0$, one uses the diffusion velocities of standard solar models. We can see that some models have radii and luminosities that are below the observed values. Thus, these models can already be rejected or at least questioned in terms of quality. The age and chemical composition intervals are completely consistent with the values obtained for the reference models of $16$CygA recalled in the third column of table \ref{tabresB}. We recall here that the models associated with ages above $7.4$ Gy were rejected for $16$CygA, based on the $t_{u}$ inversion results and their implications on microscopic diffusion and chemical composition. A successful modelling of the binary system implies finding similar ages and initial chemical composition for both stars as well as being consistent with the seismic, spectroscopic and interferometric constraints at hand. Ultimately, the models shall also be compatible with the inversion results. This is not an easy task and requires a careful analysis and a good trade-off between all of the constraints. 
\section{Inversion results}\label{SecInvRes}
In this section, we present updated inversion results for $16$CygB. In our initial work, we faced the problem of large error bars for the $t_{u}$ inversion. These error bars implied that we could not derive any additional constraints on the structure of $16$CygB. In fact the inversion results showed that all models should be accepted, regardless of whether they included diffusion or not. However, we will show in the following sections that a more careful look at the frequency data can lead to an independent diagnostic with the inversion and provide additional interesting insights on the structure of this star.

The inversion technique we present is based on the linear integral equations presented in \citet{Gough} derived for the squared isothermal sound speed $u=\frac{P}{\rho}$ and the helium mass fraction, $Y$. The basic equation of the inversion is then written:
\begin{align}
\frac{\delta \nu^{n,l}}{\nu^{n,l}}=\int_{0}^{R}K^{n,l}_{u,Y}\frac{\delta u}{u} dr + \int_{0}^{R} K^{n,l}_{Y,u} \delta Y dr, \label{EqFreqStruc}
\end{align}
where the notation $\frac{\delta x}{x}$ stands for the relative difference between observed quantities and quantities of the reference model, defined as follows:
\begin{align}
\frac{\delta x}{x}=\frac{x^{obs}-x^{ref}}{x^{ref}}. \label{eqreldiff}
\end{align}
The most striking difference between inversions in asteroseismology and inversions in helioseismology is the number of observed frequencies, leading to the fact that the classical linear kernel based inversion methods cannot be used to derive full structural profiles of observed stars. In previous studies, we have adapted the SOLA inversion techniques from \citet{Pijpers} to carry out inversions of structural integrated quantities \citep[See][for various examples.]{Reese, Buldgentu, Buldgentau}. Amongst the indicators derived, we defined a core condition indicator in \citet{Buldgentu} as follows:

\begin{align}
t_{u}=\int_{0}^{R}f(r)\left(\frac{du}{dr}\right)^{2}dr \label{EqtuRef},
\end{align}
 with $f(r)= r(r-R)^{2}\exp(-7r^{2})$, the weight function used for this inversion with $R$ the stellar radius and $r$ the radial coordinate associated with each layer inside the model, $u$ is the squared isothermal sound-speed previously defined.

First, we recall a few basic equations of seismic inversion techniques. It is important to remember that seismic diagnostics using classical inversion techniques involve individual relative frequency differences (defined as in Eq. \ref{eqreldiff}). In that sense, any inverted result is generated from a recombination of these frequency differences. When we use the linear SOLA technique \citep{Pijpers}, we build a linear combination of frequency differences. In the case of the $t_{u}$ inversion, for example, we have:
\begin{align}
\sum_{i}^{N}c_{i}\frac{\delta \nu_{i}}{\nu_{i}} \equiv \left(\frac{\delta t_{u}}{t_{u}}\right)_{inv}, \label{EqtuDefInv}
\end{align} 
with the $c_{i}$ being the inversion coefficients, which are determined by finding the optimal value of the SOLA cost function for given trade-off parameters values. The SOLA cost function is defined as follows for the $t_{u}$ indicator and denoted $\mathcal{J}_{t_{u}}$:
\begin{align}
\mathcal{J}_{t_{u}} = &\int_{0}^{1}\left[ K_{\mathrm{Avg}}-\mathcal{T}_{t_{u}}\right]^{2}dx +\beta \int_{0}^{1}K^ {2}_{\mathrm{Cross}}dx + \tan(\theta) \sum^{N}_{i}(c_{i}\sigma_{i})^{2} \nonumber \\
 &+ \eta \left[ \sum^{N}_{i}c_{i}-k \right], \label{EqCostFunc}
\end{align}
where $\mathcal{T}_{t_{u}}$ is the target function associated with the indicator, $K_{\mathrm{Avg}}$ is the averaging kernel, and $K_{\mathrm{Cross}}$ the cross-term kernel, defined with respect to the fractional radius position $x=\frac{r}{R}$. $\eta$ is a Lagrange multiplier, $k$ is a regularization factor related to the non-linear generalization of indicator inversions \citep[see][for details.]{Buldgentu}, $\sigma_{i}$ are the errors associated with each individual frequency and $\beta$ and $\theta$ are the free parameters of the SOLA method, related to the trade-off with the cross-term and the amplification of observational errors and the accuracy of the fit of the target function. Nevertheless, for this particular inversion, no additional terms used to deal with surface effects have been added since they often bias the results and reduce the quality of the fit of the target function. This is also justified by the fact that the $t_{u}$ indicator probes core regions and that its target function has low amplitude in the surface.

The averaging and cross-term kernels are defined as follows for the $(u,Y)$ structural pair, with $Y$ the helium mass fraction and $u=\frac{P}{\rho}$, the squared isothermal sound speed and the functions $K^{i}_{u,Y}$ and $K^{i}_{Y,u}$ the structural kernels associated with $u$ and $Y$ respectively:
\begin{align}
K_{\mathrm{Avg}}=\sum_{i}^{N}c_{i}K^{i}_{u,Y}, \\
K_{\mathrm{Cross}}=\sum_{i}^{N}c_{i}K^{i}_{Y,u}. \label{EqKernels}
\end{align}
The fact that we have two free parameters in the SOLA cost function is due to the ill-posed nature of the problem and leads to the well-known trade-off problem when using inversion techniques. In this particular case, the question of the trade-off is particularly important since we have three oscillation modes in particular that have larger error bars than the all the others and two of these could sometimes see their individual frequencies fitted within their error bars.  

Another specificity of asteroseismic inversions is that they are performed with little or no knowledge of the radius of the observed target, noted $R_{tar}$. In section $2.1$ of \citet{Buldgentu}, we analysed the impact of this problem on equations of the type of Eq. \ref{EqFreqStruc}. It was then shown that the inversion implicitely scaled the observed target to the same radius as the reference model used to perform the inversion while keeping its mean density constant. This meant that the target studied by the inversion was not defined by a mass $M_{tar}$ and a radius $R_{tar}$ but was a scaled target defined by a mass $\frac{M_{tar}R^{3}_{ref}}{R^{3}_{tar}}$ and a radius $R_{ref}$.

This does not restrict the diagnostic potential of the inversion technique but means that if we want to compare results from various reference models, we need to compare values of $t_{u}/R^{6}_{tar}$ to get rid of the implicit scaling process introducing a dependency in $R_{ref}$ in the inversion process.
 
\subsection{Analysis of the error contributions}\label{SecErrAn}

\begin{figure}[t]
	\flushleft
		\includegraphics[width=9cm]{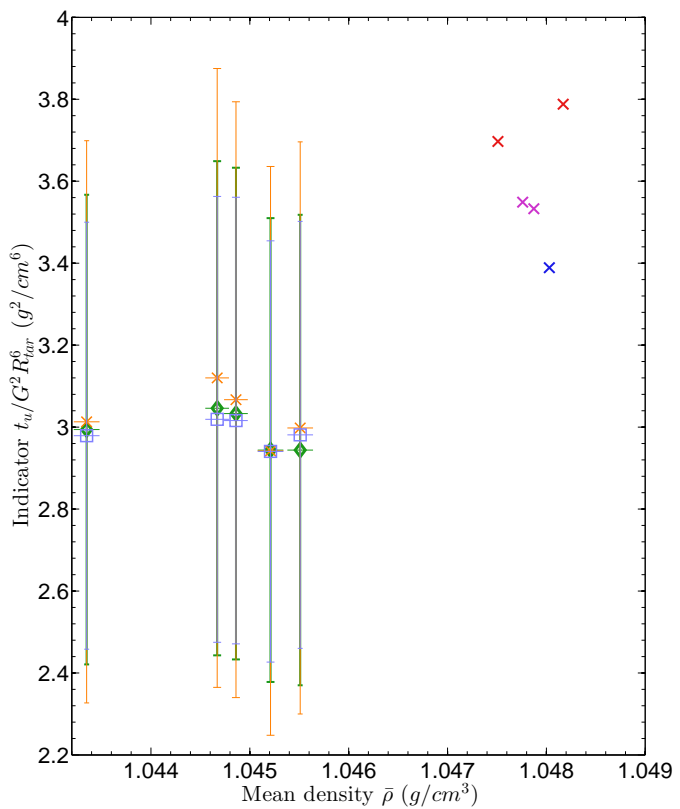}
	\caption{In orange, inversion results for the $t_{u}$ indicator and $\bar{\rho}$ with the full set of modes for $16$CygB. In green, inversion results for the same models excluding the modes with the largest error bars in the frequency set. The blue squares are associated with inversion results for which the trade-off parameter $\theta$ has been slightly enhanced. In red, blue and magenta, $t_{u}$ and $\bar{\rho}$ values in the reference models (See text for the explanation of the colour code).}
		\label{figError}
\end{figure}

In Fig. \ref{figError}, we illustrate in orange the initial inversion results of $t_{u}/R^{6}_{Tar}$ with their quite large error bars, $R_{Tar}$ being the target photospheric radius. They seemed disappointing since the kernel fits were excellent and implied that there were enough kernels to fit the target function of the $t_{u}$ inversions.

This implied that the problem was simply stemming from the observational errors propagation term in the cost function of the SOLA method. The classical way to deal with this problem is to increase the $\theta$ parameter in the cost function thus reducing the propagation of observational errors. While this may be a solution, changing the $\theta$ parameter can lead lead to a much less accurate fit of the target function and thus reduces the quality of the inversion. This implies larger errors on the inverted result coming from the kernel fit as shown in \citet{Buldgentu}. From our previous test cases, we also know that around $50$ frequencies is sufficient to obtain an inverted value for $t_{u}$, especially if octupole modes are available. Consequently, we looked at the observed frequencies for which there were large uncertainties and found that the $\ell=3$, $n=14$ mode, the $\ell=3$, $n=15$ mode and the $\ell=3$, $n=16$ had much larger uncertainties than the other modes of similar radial order. The error bars on the individual frequencies were sometimes even larger than the frequency differences between $16$CygB and the computed reference models with the Levenberg-Marquardt algorithm. This is of course somewhat inefficient since it implies that we are using frequency differences that cannot be exploited by the inversion techniques.

In fact, frequency differences with large error bars can dominate the error contribution in the inversion results, especially if the inversion coefficient associated with the particular mode is important. This is in fact simply due to the form of the term associated with the error propagation in the SOLA cost-function which is written:
\begin{align}
\sum_{i}^{N}(c_{i}\sigma_{i})^{2}. \label{EqErrorContrib}
\end{align} 

It is thus clear that modes with high inversion coefficients and large uncertainties contribute the most to the error propagation. Although the SOLA method tends to mitigate the impact of the modes with large uncertainties, the result is always a compromise between precision and accuracy. This trade-off is realized through the change of the free parameters of the inversion.In the context of asteroseismic inversions, the fact that each oscillation spectra has its own error bars, that each star is fitted individually within a given accuracy that can be variable and that each star occupies a different position in the HR diagram for which the linear approximation might be irrelevant to a certain degree, makes each inversion process unique. Therefore, from the mathematical point of view, each inversion has to be analysed differently, although trends in terms of inversion parameters can be seen and are understandable since they are linked to the data and model quality which can be objectively assessed.

The trade-off problem of inversion techniques is illustrated by the so-called trade-off curves that can be seen in the original paper on the OLA method by \citet{Backus} or \citet{Pijpers} for the SOLA method. Typically, each frequency set defines the number of coefficient available, thus the resolution of the inversion. However, this resolution is mitigated by the error bars of these individual modes which limit the amplitude of the coefficient that can be built to fit the target function. The trade-off curve materializes this competition with respect to the parameter $\theta$ of the inversion. We describe a little bit more in depth the trade-off problem and the effect of eliminating modes in the frequency spectrum in Sect. \ref{secapperror}.

Important error bars can indeed be seen for the $\ell=3$, $n=14$ mode, which is the octupole mode of lowest radial order. We know indeed from our previous test cases \citep[see][]{Buldgentu} that the $t_{u}$ inversion uses preferentially the low order modes and tends to benefit from the presence of octupole modes and use them as much as possible. Since this particular mode has the highest error bar, we wanted to see how eliminating it from the frequency set used for the inversion could help us obtain a smaller error propagation. As previously explained, inversion techniques use individual frequencies to extract information. However, this is only possible if the frequencies used by the inversions are not fitted within their observational error bars. Typically, if one eliminates a mode with large error bars, one reduces the amplification of the errors but also the resolution of the inversion. Ultimately, eliminating a mode from the frequency set is only justified if its detection is arguable or if it is already fitted within the error bars. Otherwise, reducing the error bars is more efficiently done by increasing slightly the value of the $\theta$ parameter.

In the particular case of $16$CygB, some individual modes could be fitted within their error bars and thus could not bring any additional seismic constraints if used in an inversion process. Finally, eliminating the worst offenders in terms of error bars is a process that has also been described in helioseismic inversions \citep[see][]{BasuSun}, since they can have strong impact on SOLA inversions when adjusting the trade-off parameters for the inversion.

\begin{figure*}[t]
	\flushleft
		\includegraphics[width=18cm]{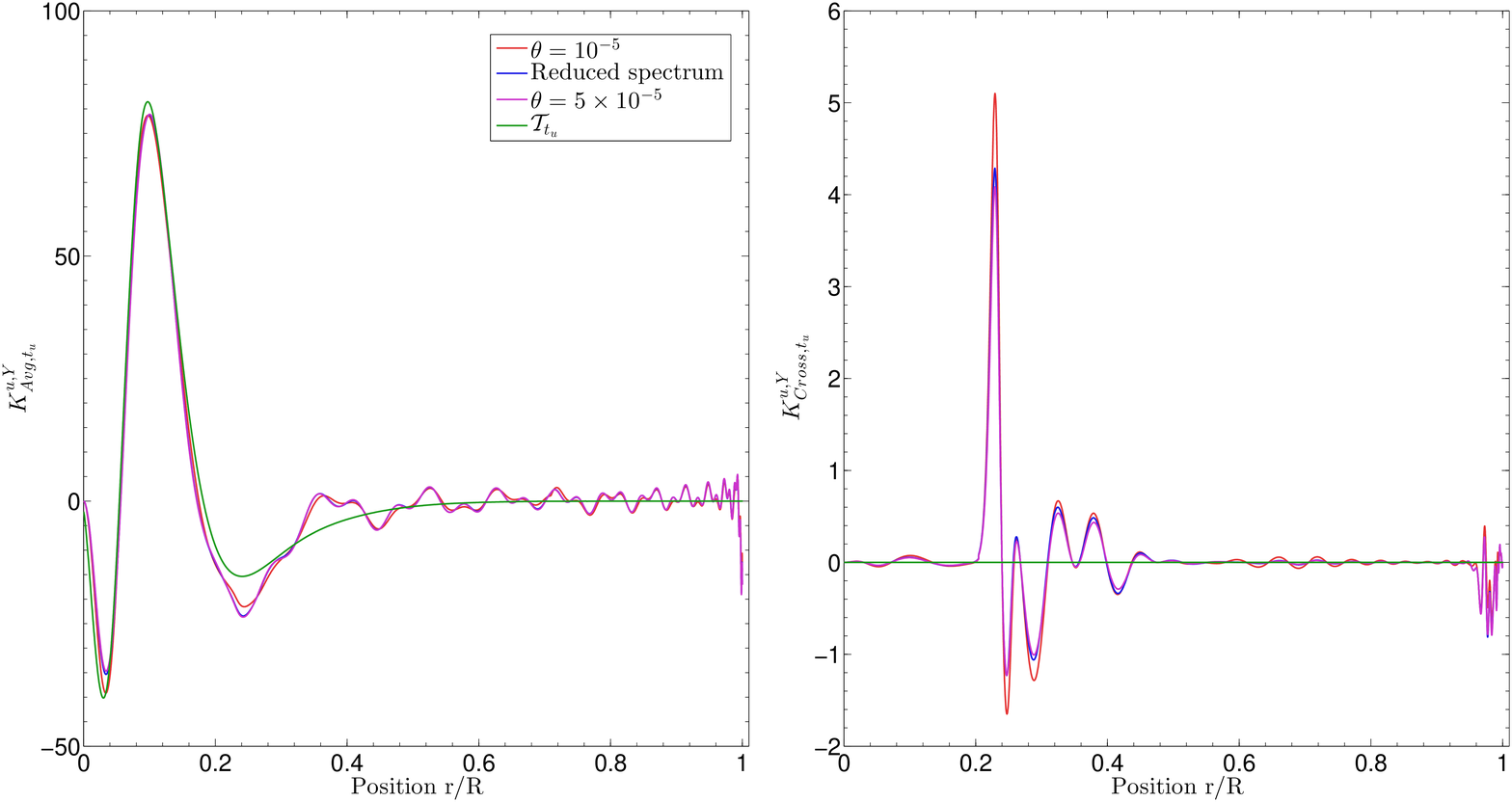}
	\caption{Left panel: averaging kernels for the core conditions indicator $(t_{u})$ for various $\theta$ values and reduced frequency spectrum. Right panel: same figure for the cross-term kernels of the $t_{u}$ inversion. We used the $(u,Y)$ structural pair in both plots.}
		\label{figKernels}
\end{figure*} 

In Fig. \ref{figError}, we show the impact of the modes with the largest error bars on the final inversion error propagation and values of $t_{u}$ and $\bar{\rho}$ for reference models. The new result with reduced error bars are illustrated in green and light-blue. The green results are obtained by eliminating problematic modes and the light-blue results are those obtained by slightly increasing the value of the $\theta$ parameter. We associated the following colour code for the values inferred from the reference models: a blue cross means that the $t_{u}$ value agrees will all inversion results, a magenta cross that it agrees with some inversion results and a red cross that it does not agree with any result. We can see that eliminating the modes with large uncertainties reduces significantly the error bars on the inverted result, without changing much the actual inverted value. A change in the inverted value would have meant that the mode had a significant impact on the inversion result. In practice, this could be seen by a change in the fit of the target function by the averaging kernel. This could be the case if one had fewer individual frequencies and that the problematic oscillation mode was used by the inversion despite its large error bars. In figure \ref{figKernels}, we illustrate the change in the averaging and cross-term kernel fit that is induced by the elimination of the most problematic modes in terms of observational error bars and an increase of the trade-off parameter $\theta$. As was the case for the inverted $t_{u}$ values, the differences on the averaging kernels are minimal. Hence, an independent study of $16$CygB in terms of $t_{u}$ can be performed. In the next section, we present new inversions results using a greater number of models for different surface chemical compositions, yet within the observational constraints, and for different diffusion coefficients, in much the same way as what was done in our previous study, more focused on $16$CygA.
\subsection{$t_{u}$ Inversion for $16$CygB}\label{SecRestu}
In this section, we present the results for the $t_{u}$ inversions for $16$CygB. Using the reference models computed with our Levenberg-Marquardt algorithm and the more regularized inversions, we were able to obtain a value of $t_{u}$ for $16$CygB with lower error bars. However, the uncertainties are still non-negligible. Thus, we have to combine our analysis with other diagnostics and carefully discuss our final results, as was done in our previous study of $16$CygA. In Fig \ref{figBox}, we present our results for various models with various surface chemical composition and changes of the factor $D$ associated with atomic diffusion in CLES. The results we obtain are slightly model-dependent, which is very similar to what was obtained for $16$CygA, but the trend is in this particular case opposite to what was seen before. Indeed, in \citet{BuldgenCyg}, we saw that including microscopic diffusion provided much more consistent values of the $t_{u}$ indicator when compared to the inverted values. For $16$CygB, models with lower helium surface abundances, higher surface metallicities and less diffusion are favoured. In fact, reducing the $t_{u}$ value is directly related to a reduction of the gradient of $u=\frac{P}{\rho} \approx \frac{T}{\mu}$, with $T$ the temperature and $\mu$ the mean molecular weight. Consequently, reducing $t_{u}$ implies reducing the mean molecular weight gradient within the star or changing the temperature gradient in the regions where the $t_{u}$ indicator is sensitive. Reducing the mean molecular weight gradient can first be done by eliminating microscopic diffusion in the models. Indeed, this process tends to accumulate heavy elements in the deeper regions since for stars around $1.0M_{\odot}$, gravitational settling dominate the transport mechanism in the deep radiative regions. However, as stated before, not including this process leads to inconsistent ages and chemical compositions for both stars. Therefore, the reason for this discrepancy has to be explained using a more subtle effect.

\begin{figure*}[t]
	\flushleft
		\includegraphics[width=17cm]{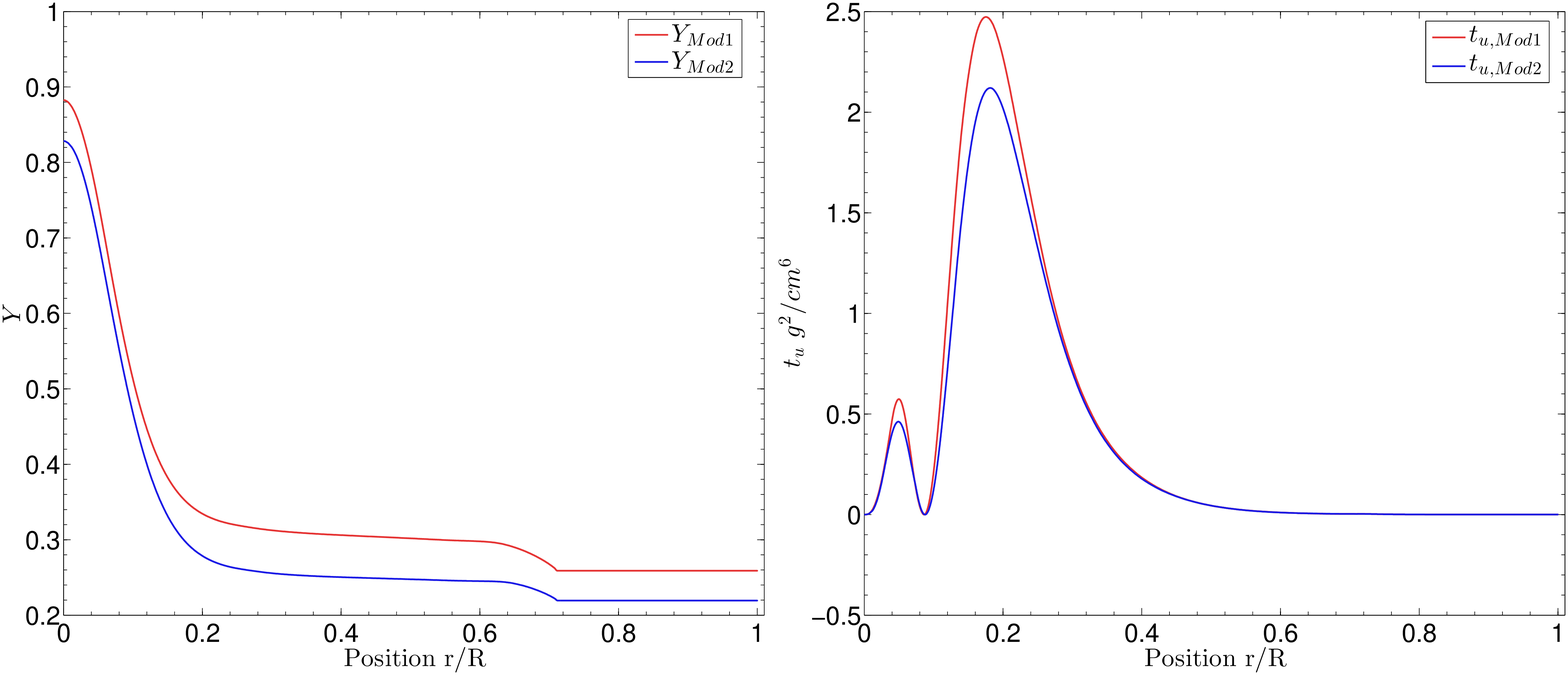}
	\caption{Left panel: In blue, helium abundance profile $(Y)$ for one model with a lower surface helium abundance, around $0.22$. In red, $Y$ profile for a model with a higher surface abundance, around $0.26$. Right panel: the profile of the target function of the core conditions indicator $(t_{u})$ is plotted in corresponding colours for both models.}
		\label{figTuY}
\end{figure*}

We show in Fig. \ref{figTuY} the differences in chemical composition and in the weight function involved in the integral expression for the $t_{u}$ indicator for two of our reference models in the chemical composition box. Model$_{1}$ is a model with a higher helium content $(Y_{S}=0.26)$, lower metallicity $((Z/X)_{S}=0.0208)$ and microscopic diffusion $(D=1.0)$, thus following the prescription derived from our previous study. Consequently, it is also less massive $(M=0.91 M_{\odot})$ and within the ``young'' range of our reference models $(Age=7.32Gy)$. Due to the higher helium content and efficient microscopic diffusion, this model is rejected by the $t_{u}$ inversion. Model$_{2}$ has a low helium content $(Y_{S}=0.22)$ and a higher metallicity $((Z/X)_{S}=0.0214)$ and a less efficient microscopic diffusion $(D=0.5)$. This model is significantly more massive than Model$_{1}$ $(M=1.01M_{\odot})$ but has a quite similar age of $7.54Gy$. The strong difference in mass is due to the well known degeneracy associated with the helium abundance. It should be noted that this model is validated by the $t_{u}$ inversion.

This illustrates the fact that simply changing the surface chemical composition or microscopic diffusion has a strong impact on the fundamental parameters of the star and implies strong changes in the internal structure even if the model fits all the observational constraints (although Model$_{1}$ should be rejected due to its lower radius). Both models were chosen because they were extreme cases and illustrated well the strong degeneracy due to helium abundance. 

\subsection{Comparison with $16$CygA}

If we consider again $16$CygA, the models with masses around $1.01M_{\odot}$, high helium content and ages $7.2$Gy around were considered to be the best models of this star since they reproduced the $t_{u}$ trend seen in our previous paper. This would mean that we would chose a model closer to Model$_{1}$ to be consistent in terms of the initial chemical composition of both components. However, since in this case we have to reduce the $t_{u}$ values, and thus apply the opposite changes to the chemical composition and microscopic diffusion, the $16$CygB have higher masses and ages (like Model$_{2}$ mentionned above), going up to $1.03M_{\odot}$ and $8.0$Gy.

The fact that the inversion is able to distinguish between Model$_{1}$ and Model$_{2}$ proves again the diagnostic potential of this approach. In this particular case, due to the fact that both stars are within a binary system, we are even able to see whether our selected result will be consistent with the previously determined parameters for $16$CygA. Due to the very similar chemical composition derived spectroscopically and seismically, due to the results of independent forward modelling of both components leading to similar ages and initial chemical composition, we rather consider that the differences seen with the inversion technique is to be explained by inaccuracies in the models rather than considering the binary system to have merged from two isolated stars. 

\begin{figure*}[t]
	\flushleft
		\includegraphics[width=18cm]{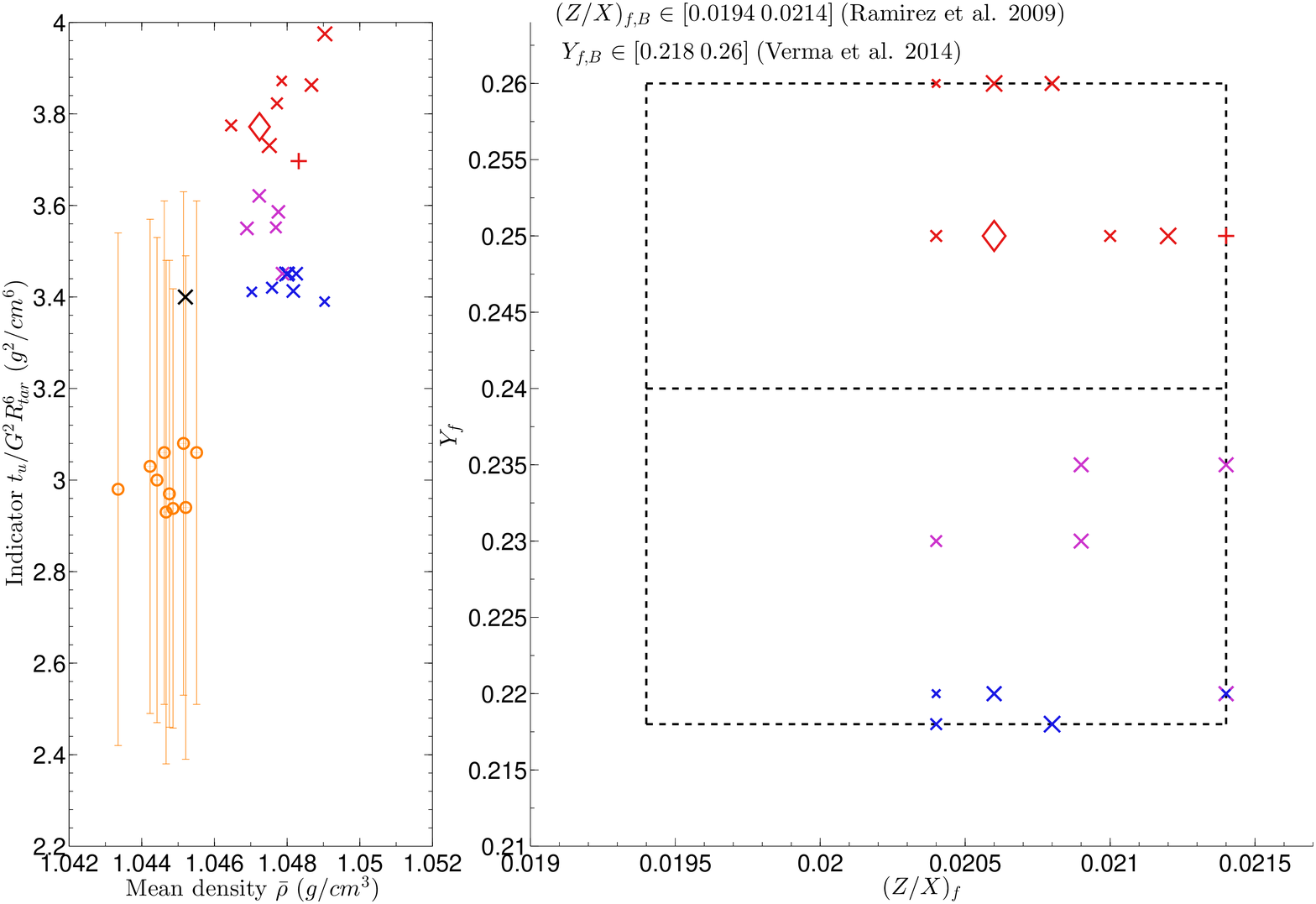}
	\caption{Left panel: mean density $(\bar{\rho})$ vs core conditions indicator $(t_{u})$ plot. The inversion results are plotted in orange with their respective error bars. The crosses are values for the reference models computed with the Levenberg-Marquardt algorithm with AGSS$09$, the black $\times$ shows one example of a model computed with GN$93$ and $Y_{f}=0.25$. Right panel: surface chemical composition box for $16$CygB. The colour code used allows direct trend comparisons between the surface chemical composition and the $t_{u}$ values. The size of the symbols is related to the intensity of microscopic diffusion, the smaller the symbol, the smaller the $D$ coefficient. The $+$ and the $\lozenge$ illustrate the impact of the metallicity on the $t_{u}$ value.}
		\label{figBox}
\end{figure*}

We also illustrate in Fig. \ref{figBox} the results for one model using the GN$93$ abundances and models which were computed using the AGSS$09$ abundances and assuming a similar initial chemical composition to what was derived for $16$CygA in our previous study. These models show values of $t_{u}/R^{6}_{tar}$ around $3.7$ $g^{2}/cm^{6}$ whereas the model with GN$93$ is more consistent with the inversion results of $3.0\pm0.5$ $g^{2}/cm^{6}$. It is clear that models computed assuming the same ingredients as $16$CygA are incompatible from the point of view of the inversion. However, since these stars form a binary system and thus are thought to have formed together, we should be able to derive similar values of the initial chemical abundance and similar ages for both components of the system. This problem is also reflected in the effective temperature and radii determination. The well-known helium mass degeneracy leads to smaller radii for models with higher helium abundances, for example. We also tried using larger error bars on the effective temperature and looked at models with $T_{eff}$ between $5600$ $K$ and $5900$ $K$ to see if this could affect the results. Ultimately, no trend was found since they are ultimately related to the chemical abundances and the way the elements are mixed within the star. These effects are well-known to affect the position of the models in the HR diagram at the end of its evolution. Thus, in what follows, we will focus on these aspects to try to reconcile our models of $16$CygB with those of $16$CygA and the inverted results.

In terms of precision and accuracy, it should be noted that neither the model-dependency, nor the regularisation can be held responsible for an inaccurate result. Hence, as shown in this section, in particular thanks to the large variety of reference models, we can see that surface chemical composition changes are not sufficient to explain the inverted values of $t_{u}$. In fact, taking $\theta=10^{-4}$ still implies very similar inversion results with reduced error bars and a slightly worse fit to the target function. Moreover, we know from our previous numerous test cases that the $t_{u}$ inversion provided accurate seismic diagnostic of core regions \citep[See][]{Buldgentu}.

\subsection{Influence of physical parameters on $t_{u}$}

When analysing the effects of microscopic diffusion, the problem is even worse, since if we trust the values of $Y_{f} \in \left[ 0.24\; 0.25 \right]$ for the final surface helium abundance of $16$CygA, we should obtain higher $Y_{f}$ values for its less-massive counterpart due to the fact that its convective envelope goes slightly deeper and implies less-efficient microscopic diffusion. One should note that similar conclusions can be drawn for the surface heavy element abundance of this star. In fact, increasing the amount of heavy elements in the stars increases the opacity in the deep radiative regions where the $t_{u}$ indicator is sensitive (see Fig. \ref{figKernels}). Thus, it implies an increase of the temperature gradient, $\frac{dT}{dr}$. Now, since $t_{u}\propto \left(\frac{du}{dr} \right)^{2}$ (see Eq. \ref{EqtuRef}), it is worth looking more in depth at the behaviour of this indicator with changes in the stellar structure. Using the ideal gaz approximation, we have a straightforward relation between $u$, $T$ and $\mu$.

\begin{align}
\left(\frac{du}{dr}\right)^{2}&\approx \frac{T^{2}}{\mu^{2}}\left(\frac{d \ln T}{dr}-\frac{d \ln \mu}{dr}\right)^{2}. \label{EqGrad}
\end{align}

\begin{figure*}[t]
	\flushleft
		\includegraphics[width=17cm]{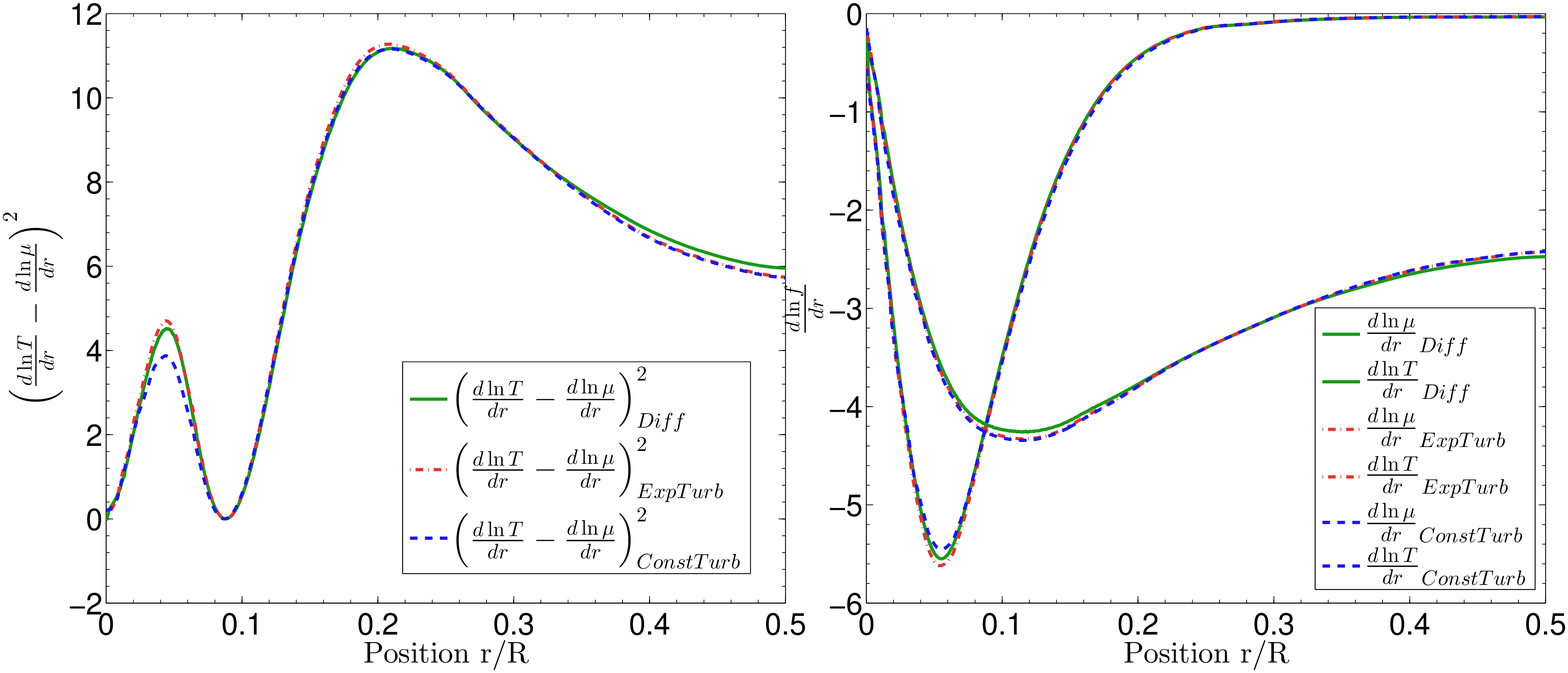}
	\caption{Left panel: plot showing the difference of the gradient of the natural logarithm of temperature $(T)$ and that of the mean molecular weight $(\mu)$ for models including different mixing processes: the green curve is for a model with microscopic diffusion, the blue curve is for a model with a constant turbulent diffusion coefficient and the red curve is for an exponentially decaying turbulent diffusion coefficient. Right panel: the gradient of the natural logarithm of the mean molecular weight and of the temperature for the same models as in the left panel, the colour code has been respected.}
		\label{figStruc}
\end{figure*} 
\begin{figure*}[t]
	\flushleft
		\includegraphics[width=17cm]{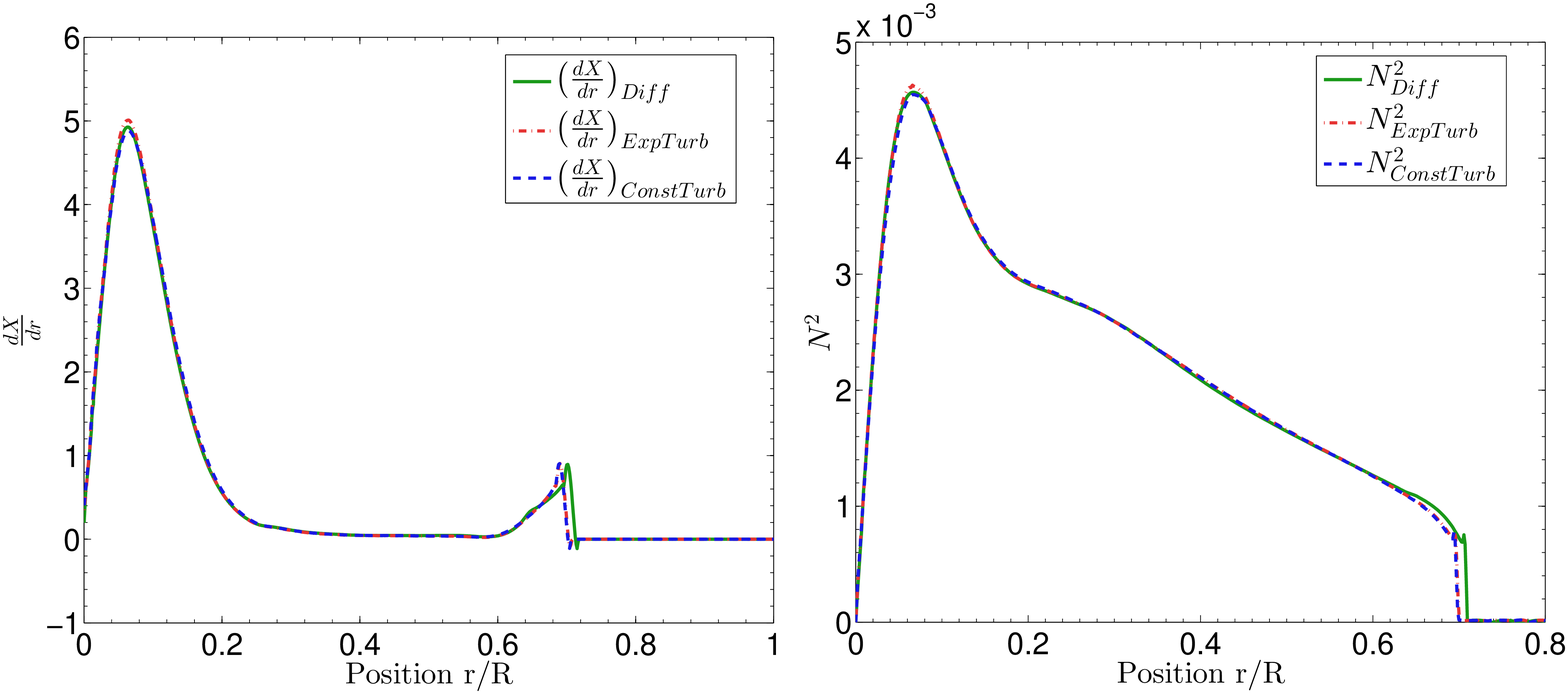}
	\caption{Left panel: plot showing the hydrogen gradient of the same models as in Fig. \ref{figStruc}. Left panel: plot showing the Brunt-Väisälä frequency of these models.}
		\label{figStruc2}
\end{figure*} 
This formula implies that the behaviour of the indicator depends on the values of the gradients themselves. As can be seen in the right panel of Fig. \ref{figStruc}, it is not always straightforward to say whether an increase of the mean molecular weight gradient through diffusion will imply an increase of $t_{u}$. For example, below $0.1$ R, with $R$ the stellar radius, it will be the case because diffusion will increase the depth of the minimum just below $0.1$ R. However, adding extra mixing around $0.2$ R or $0.3$ R will smooth the transition towards the chemically homogeneous convective envelope (around $0.7$ for this model) thereby decreasing the value of $t_{u}$. Similarly, increasing the temperature gradient below $0.1$ will reduce $t_{u}$, and reducing $\frac{dT}{dr}$ above $0.1$ (thus sharpening the transition towards the convective envelope) will imply the same reduction for the indicator. One can see these effects in Fig. \ref{figStruc} where we illustrated the impact of different types of mixing on the temperature and $\mu$ gradients and thus on the $t_{u}$ indicator. This gives us a clue as to what could be modified in the models to reconcile the inversion results with the other constraints. However, it does not mean that this is the only solution to the problem we presented previously. For the sake of illustration, we also illustrate the hydrogen gradient and the Brunt-Väisälä frequency of these models, showing the change of the slope of the hydrogen gradient at the bottom of the convective zone but also a significant deplacement of the base of the convective zone for these models due to the use of the new opacities from the OPAS project. The OPAS opacities are new opacity tables specifically designed for solar-like conditions, where great care has been given to the details of the absorption lines considered. These models also used the latest version of the OPAL equation of state \citep{Rogerseos}. These changes of course affect the stratification below the convective zone and thus the behaviour of the $t_{u}$ indicator. Turbulent diffusion also implied a change of the Brunt-Väisälä frequency in the very deep regions (below $0.1$ R), this is particulary seen for the model associated with constant turbulent diffusion.

In practice, all thermodynamic quantities are coupled through the equation of state. For example, adding a mixing process will affect the chemical composition, thus the mean molecular weight, but it will also affect the opacity and indirectly the temperature gradient. Consequently, the $t_{u}$ inversion offers a new insight on some differences between the target and the reference model, but does not provide the physical cause of the observed differences in structure.
\section{Impact of physical ingredients on the core conditions indicator}\label{SecDisc}
\subsection{Adding extra mixing}
\begin{figure*}[t]
	\flushleft
		\includegraphics[width=18cm]{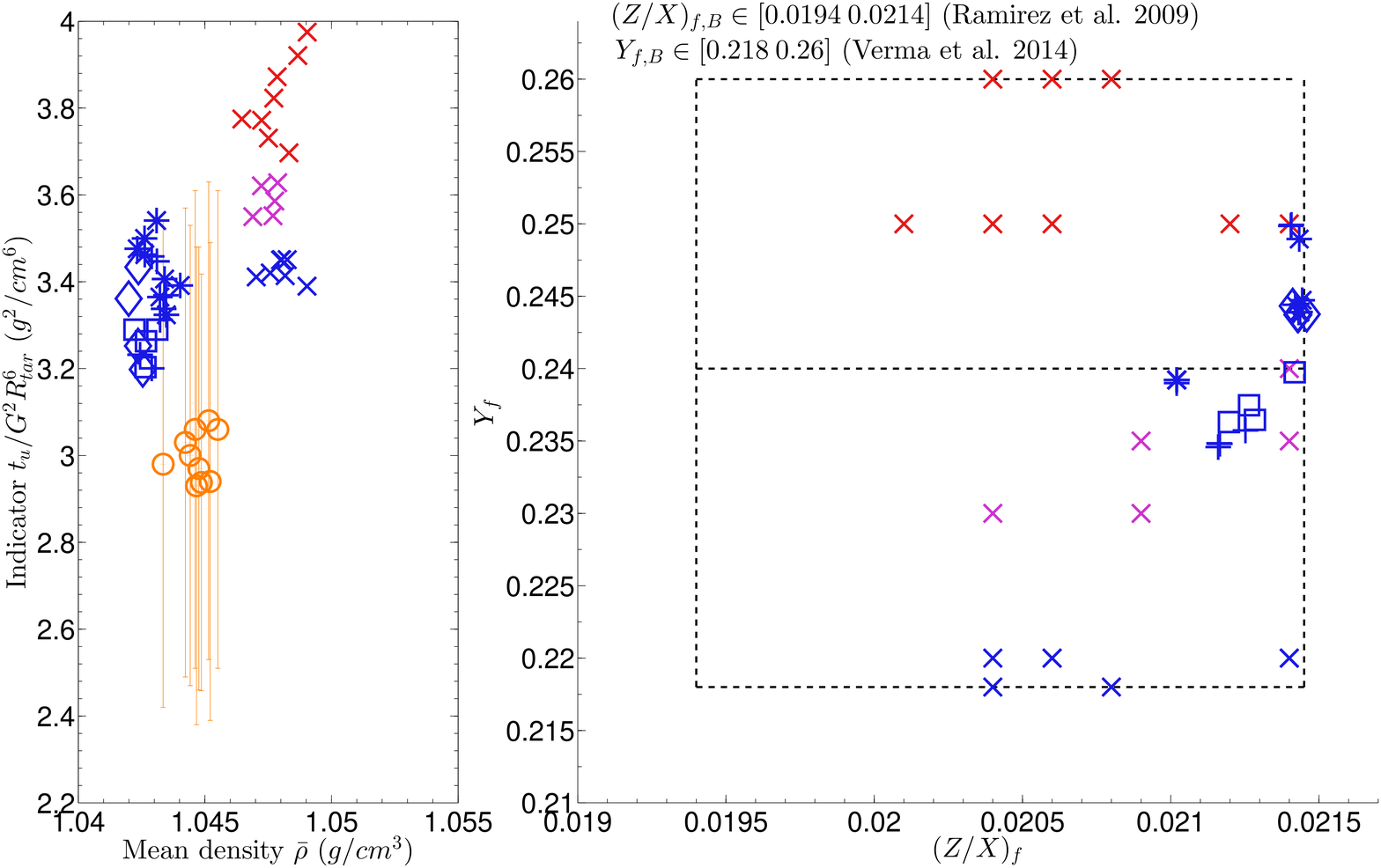}
	\caption{Left panel: mean density $(\bar{\rho})$  vs core conditions indicator $(t_{u})$ plot. The inversion results are plotted in orange with their respective error bars. The $\times$ are values for the reference models computed with the Levenberg-Marquardt algorithm without extra mixing. The $\square$ are related to models with a decaying exponential turbulent diffusion coefficient starting at the bottom of the convective enveloppe. The $+$ show models with a decaying exponential diffusion coefficient starting from the surface and the $\lozenge$ use a constant turbulent diffusion coefficient. The $*$ depict models using the new OPAS opacities and the decaying exponential coefficient starting from the surface. Right panel: surface chemical composition box for $16$CygB. The colour code allows direct trend comparisons between the surface chemical composition and the $t_{u}$ and $\bar{\rho}$ values as in Fig. \ref{figBox}.}
		\label{figBoxMix}
\end{figure*}

Because of the $t_{u}$ inversion results, we are faced with a very peculiar problem. We have two stars, in a binary system, with very similar surface chemical composition, similar masses and radii, that show significantly different seismic behaviours when carrying out inversions of their structure. The problem is that the models for both stars cannot be consistent with the inversion results and simultaneously present similar chemical composition and age. Small discrepancies in chemical composition between both stars have proven not to be sufficient to eliminate the discrepancy with the inverted $t_{u}$ values. Therefore, we had to assume that something was neglected in the models for $16$CygB, or $16$CygA, or for both stars. In what follows, we study supplementary models including a parametrized approach for an additional mixing process. The physical nature of this mixing process is not discussed here, but we demonstrate that the $t_{u}$ indicator is, as expected, able to discriminate between various processes inside the star. Figure \ref{figBoxMix} shows various $t_{u}$ inversion results for different implementations of diffusion yielding different chemical compositions. At first, we still wish to see whether there is a way reconcile the chemical composition of $16$CygB with that of $16$CygA.

The parametrization of this additional mixing is based on an implementation of turbulent diffusion used in previous studies \citep[see][for details]{Miglio}. We tested different implementations of this mixing. First, we added a constant turbulent diffusion coefficient of around $20cm^{2}s^{-1}$ acting in the entire stellar structure and computed a few models fitting the observational constraints for $16$CygB. The impact of the constant turbulent diffusion coefficient is quite strong. Indeed, gradients are quickly attenuated and the $t_{u}$ value decreases, as can be seen in Fig. \ref{figBoxMix} with the positions of the blue $\lozenge$ in the left panel. However, disagreement with other constraints is quickly found if this mixing is further increased. For example, it is impossible to fit the individual small frequency separations when the extra mixing is too important although the acoustic radius, the mean density and other constraints of the cost function of the forward modelling can be accurately fitted.

We also computed models with a diffusion coefficient implemented as an exponential decay starting either from the bottom of the convective envelope or from the surface. Two parameters are used for this formalism, one multiplicative constant and the rate of exponential decay. From previous studies \citep{Miglio}, we know that a multiplicative coefficient of around $100cm^{2}s^{-1}$ is consistent with the effects of rotation expected in solar-like stars. This value was used as a benchmark for the order of magnitude of the mixing, but we did not limit ourselves to this value since we wanted to investigate the effects of this parametric implementation on the $t_{u}$ indicator. We thus allowed changes of up to $\pm 50cm^{2}s^{-1}$ in the value of this diffusion coefficient. From Fig. \ref{figBoxMix}, where the models with the implementation of turbulent diffusion as an exponential decay starting from the surface are represented by blue $+$, we can see that it can indeed help to reconcile the models with the $t_{u}$ values for $16$CygB, even if a higher present surface helium value is considered, as had to be done for $16$CygA. The fundamental parameters of these models are presented in table \ref{tabNewresB}, we note that they have slightly higher masses and ages than the models without turbulent diffusion for the same chemical composition. 

\begin{figure*}[t]
	\flushleft
		\includegraphics[width=18cm]{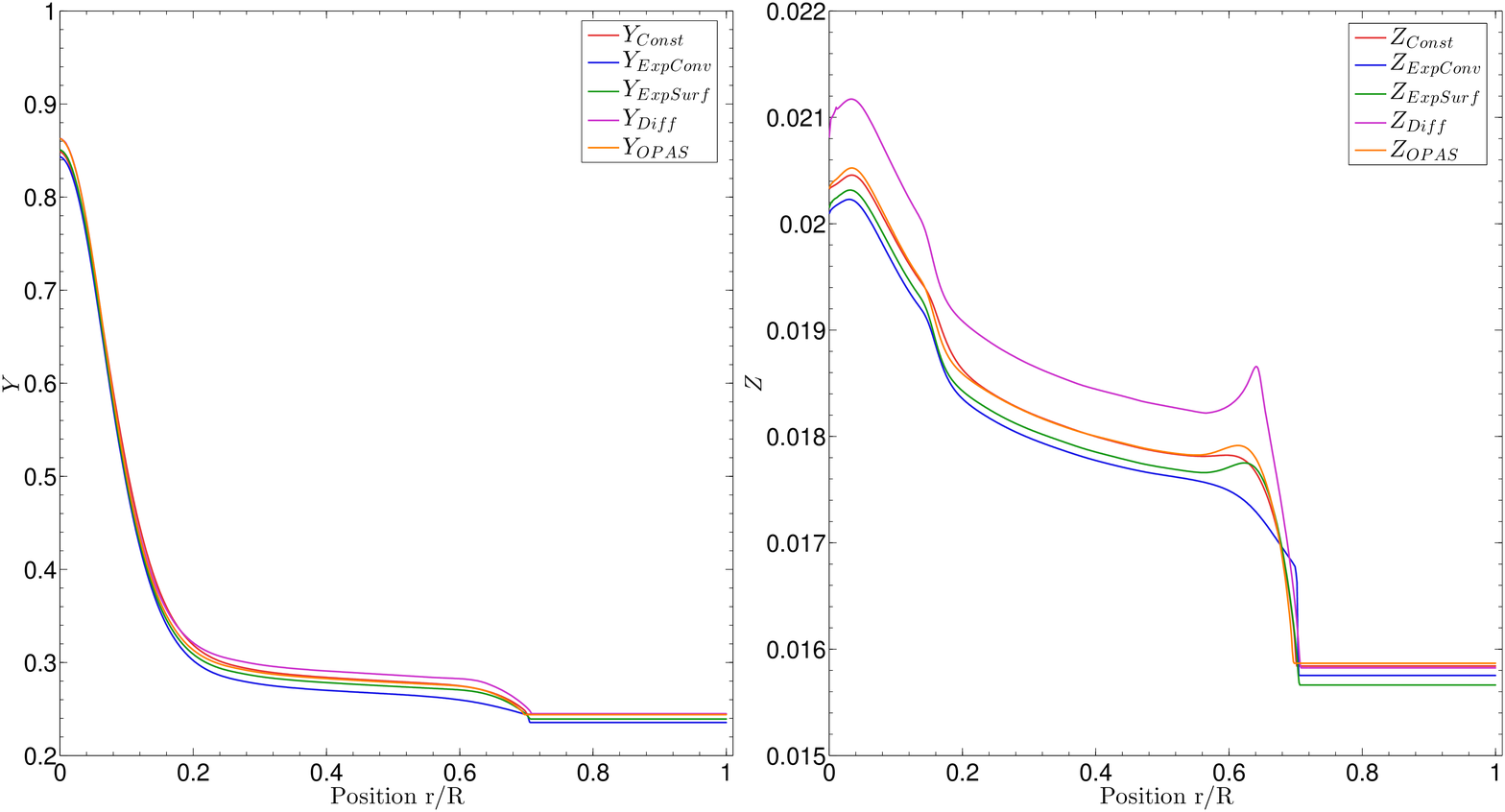}
	\caption{Left panel: helium abundance plot for three models with different implementations of turbulent diffusion. For the red curve, a constant mixing coefficient was applied throughout the structure and the evolution of the model. For the blue curve, we used an exponential decay starting from the base of the convective envelope of the model. For the green curve, we used an exponential decay starting from the surface of the model. For the magenta curve, we used a model including only microscopic diffusion. Right panel: The metallicity profiles of the three models described above.}
		\label{figProfMix}
\end{figure*}

As expected, additional mixing can indeed help to reconcile the chemical compositions of both stars, but does not reconcile them in age since some of the models computed with the extra mixing have ages up to $8$Gy even if most are still around $7.4-7.7$Gy. It is also noticeable that the masses of the models in the present study tend to be slightly higher than those previously found in \citet{BuldgenCyg} using constraints from the $16$CygA modelling. However, all of these models are still consistent with the radius, luminosity and $\log g$ constraints from the litterature. A clear trend is also seen in the fact that increasing the mixing improves the agreement between the reference models and the inversion. However, as the models come closer to the inverted values for $t_{u}$, they tend to be less consistent with the small frequency separation values, meaning that the extra mixing should not be too intense. Indeed, reducing the rate of exponential decay (thereby extending the effects of extra mixing to lower regions) or increasing directly the turbulent diffusion coefficient leads to the same disagreement with the small frequency separations. To better understand the problems here, we plot the effects of the extra mixing on both the metallicity and helium profiles in Fig. \ref{figProfMix}. We see that the main effect is to reduce a metallicity peak right under the convective region. The more reduced the peak is, the closer the $t_{u}$ values to the inverted ones. But in the meantime, we also degrade the agreement with the small frequency separations. Changes are also seen for the helium profile right under the convection zone. During the fitting process with the Levenberg-Marquardt algorithm, this affects the initial helium abundance required to be within the constraints from \citet{Verma} and thus indirectly the hydrogen profile and thus the metallicity constraint.

\subsection{Changing the opacities}
To further investigate the problem, we computed additional models with the new OPAS opacities \citep{Mondet,Lepennec} including atomic diffusion and the implementation of turbulent diffusion using a exponential decaying function starting from the surface. We used our forward modelling approach to compute these models and analysed whether they models could agree better with both the small frequency separations and the inverted values of $t_{u}$. As stated before, adding extra mixing could reduce the agreement with the small frequency separations if its intensity was too high. However, when using the new opacities, we were able to further increase the intensity of the extra mixing, and thus the agreement with the $t_{u}$ inversion, without degrading the agreement with the small frequency separations. As such, they provide a partial help to the problem of fitting the all constraints, as can be seen from the position of the blue $*$ in Fig. \ref{figBoxMix}, but do not solve completely the problem. We can see the influence of these increased opacities, slightly moving deeper the base of the convective envelope and changing the efficiency of the mixing right below the base of the convective zone. We recall here that these models were still selected with the Levenberg-Marquardt algorithm using the observational constraints of $16$CygB. The fundamental parameters of these models computed with were very similar to those obtained previously, but these models tended to show a slightly lower luminosity around $1.18L_{\odot}$ and mass around $0.98M_{\odot}$ and are thus responsible for the "lower" part of the intervals given in table \ref{tabNewresB}.

\begin{table*}[t]
\caption{Parameters of the models of $16$CygB with extra mixing and OPAS opacities}
\label{tabNewresB}
  \centering
\begin{tabular}{r | c | c }
\hline \hline
 & \textbf{ $16$CygB models} & \textbf{ $16$CygB models } \\
 & \textbf{(Mixing)} & \textbf{(OPAS + mixing)}\\ \hline
 \textit{Mass ($\mathrm{M_{\odot}}$)}& $0.98$-$1.00$ & $0.96$-$0.99$\\
\textit{Radius ($\mathrm{R_{\odot}}$)}& $1.07$-$1.10$ & $1.07$-$1.09$\\ 
\textit{Age ($\mathrm{Gyr}$)} &$7.2$-$7.6$ & $7.3$-$7.5$\\ 
\textit{$L_{\odot}$ $(\mathrm{L_{\odot}})$} & $1.19$-$1.22$ & $1.17$-$1.20$\\
\textit{$Z_{0}$} & $0.0180$-$0.0190$ & $0.185$-$0.019$\\
\textit{$Y_{0}$} & $0.28$-$0.30$ & $0.28-0.30$\\
\textit{$\alpha_{\mathrm{MLT}}$} & $1.78$-$1.90$ & $1.75$-$1.8$\\
\hline
\end{tabular}
\end{table*}

It can also be seen that when using a turbulent diffusion coefficient decaying from the lower boundary of the convective region, the effect on $t_{u}$ is slightly more efficient, as illustrated by the position of the blue $\square$ in the $\bar{\rho}-t_{u}$ diagram of Fig. \ref{figBoxMix}. Nevertheless, we did not seek here to fine-tune the parameters in this study since we are using a parametric approach to the problem without any physical background.

At this stage, we can already conclude that reconciling both models in terms of chemical composition and age will also probably need to remodel $16$CygA to analyse whether effects other than diffusion could not be held responsible for the trend in $t_{u}$ previously observed. In that sense, looking at constraints from the lithium abundance \citep{King} and combining these constraints in the modelling of both stars might change the derived fundamental parameters by a few percents. 
\section{Conclusion}\label{SecConc}
In this paper, we updated our study of the $16$Cyg binary system by focusing our attention on $16$CygB. From a re-analysis of the data, we were able to extract information from the $t_{u}$ inversion and analyse the impact of extra mixing on the $t_{u}$ values and other classical seismic indicators. First, we illustrated and solved the problem associated with the propagation of observational errors for inversions in $16$CygB by analysing the impact of trade-off parameters and the presence of modes which in some cases were useless for the inversion technique. Ultimately, this approach could be used in similar situations for other observed targets.

From the $t_{u}$ inversion, we were able to expose a problem in the surface chemical composition of $16$CygB when compared to its companion. We computed a new set of models for this star, varying the surface chemical composition and restricting the effect of diffusion. We then observed that when the models were consistent with the inversion results, they were systematically inconsistent with the surface chemical composition we obtained for $16$CygA. Since changing the chemical composition was not the solution, we sought to implement an extra mixing process in the models of $16$CygB and tried to analyse its potential impact on the $t_{u}$ values. As intuitively guessed, an extra mixing in the form of turbulent diffusion was found to be able to reconcile the models both with the surface chemical composition of $16$CygA and the inversion results. Furthermore, using the new OPAS opacity tables further improved the agreement with the inversion. One could argue that other implementations could be tested, such as extra mixing in the form of undershooting using the prescription of \citet{Zahn} as was found in HD $52265$ by \citet{Lebreton}. However, as was described in \citet{Lebreton}, this extra mixing would leave an oscillatory pattern in the $rr_{01}$ and $rr_{10}$ seismic indicators. Due to the quality of the seismic data of $16$CygB, we were able compute these indicators and found no evidence for an oscillatory pattern but rather a decreasing trend with frequency that is well reproduced by models without undershooting.

To conclude, we can state that various physical processes could improve the agreement. For example, a change in opacity would further change the results of the forward modelling process and thus the stellar parameters obtained with the Levenberg-Marquardt algorithm, these new models could potentially be in agreement with the inversion of the $t_{u}$ indicator. In this study, extra mixing in the form of turbulent diffusion was invoked to reduce the disagreement between the inversion and the models. However, we did not seek to provide a physical explanation for this mixing and while it helps reducing the disagreement, further studies need to be performed to completely solve the problem. One first point would be to re-analyse $16$CygA in the scope of the impact of extra mixing. Indeed, we have shown here that turbulent diffusion can change the $t_{u}$ values. It is also well-known that rotation induces such type of extra mixing and it is believed to be responsible for the destruction of lithium in stars. Therefore, a first step would be to perform a thorough study of the impact of extra mixing on lithium abundances and inversion results for $16$CygA. The case of $16$CygB should be re-analysed afterwards, since it is well-known that this star shows even lower lithium abundances and is believed to have triggered thermohaline diffusion by accreting planetary matter \citep[See][]{Deal}. As such, combining spectroscopic and seismic constraints in this binary system may provide new insights on stellar modelling of solar-like stars.

Moreover, additional indicators obtained through inversions seem to be a promising way to analyse the boundaries of convective envelopes. Consequently, from the sensitivity of seismic inversions and the quality of additional constraints, we are convinced that a re-analysis of the $16$Cyg binary system with new stellar models should shed new lights on extra mixing processes in stellar interiors.

\begin{acknowledgements}
G.B. is supported by the FNRS (``Fonds National de la Recherche Scientifique'') through a FRIA (``Fonds pour la Formation à la Recherche dans l'Industrie et l'Agriculture'') doctoral fellowship. This article made use of an adapted version of InversionKit, a software developed in the context of the HELAS and SPACEINN networks, funded by the European Commissions's Sixth and Seventh Framework Programmes.
\end{acknowledgements}
\bibliography{biblioarticle4}
\appendix
\section{The trade-off problem for $t_{u}$ inversions}\label{secapperror}
\subsection{Origin of the trade-off problem}
As described in section \ref{SecInvRes}, the SOLA inversion technique we use to obtain values of the $t_{u}$ indicator computes a linear combination of individual frequency differences. These coefficients are obtained through the minimization of the cost-function defined as Eq. \ref{EqCostFunc}. We recall this definition here to better analyse the different contributions:
\begin{align}
\mathcal{J}_{t_{u}} = &\int_{0}^{1}\left[ K_{\mathrm{Avg}}-\mathcal{T}_{t_{u}}\right]^{2}dx +\beta \int_{0}^{1}K^ {2}_{\mathrm{Cross}}dx + \tan(\theta) \sum^{N}_{i}(c_{i}\sigma_{i})^{2} \nonumber \\
 &+ \eta \left[ \sum^{N}_{i}c_{i}-k \right], \label{EqCostFuncErr}
\end{align}
The first integral is associated with the averaging kernel, denoted $K_{\mathrm{Avg}}$, this term defines the accuracy of the inversion technique, the better the fit of the target function, here $\mathcal{T_{t_{u}}}$, the more accurate the inversion is.

The second integral is associated with the cross-term kernel, denoted $K_{\mathrm{Cross}}$. The cross-term stems from the presence of a second integral in Eq. \ref{EqFreqStruc}, here for example associated with $Y$. Since the inversion only wants to extract information from the variable $u$, the contribution associated with $Y$ as here to be damped. The trade-off between the reduction of the cross-term and the fit of the target function is calibrated by the free parameter $\beta$. In the case of the $t_{u}$ inversion, the use of $Y$ means that cross-term kernels have naturally smaller amplitudes due to the intrinsic small amplitude of helium kernels, as can be seen by comparing the right and left panels of Fig. \ref{figKernels}. From previous hare and hounds, we know that the cross-term contribution is much smaller then the errors from the averaging kernels and have a negligible effect on the inversion results.

The third term of Eq. \ref{EqCostFuncErr} is associated with the propagation of observational errors. This term regulates the precision of the inversion technique by damping the coefficients associated with large error bars. Since large coefficients are required for the $t_{u}$ inversion, this term has an important impact on the final outcome of the inversion is at the centre of the trade-off problem we will discuss. The importance given to the observational error bars of individual oscillation mode is materialized by the free parameter $\theta$. Ultimately, the SOLA method comes down to a trade-off between precision and accuracy. In practice, a large value of $\theta$ will imply small error bars, but also potentially a very bad fit of the target function and the reduction its accuracy. On the opposite, a very small value of $\theta$ means that the target function is well-fitted, but the result cannot be trusted due to its large error bars.

The fourth term is associated with an additional regularization based on homologous relations. The proper justification of the value of the coefficient $k$ can be found in Sect. $3.2$ of \citet{Buldgentu} and additional examples can be found in \citet{Reese} and \citet{Buldgentau} for other indicators. Eta is thus no free parameter but a Lagrange multiplier.
\subsection{Effects of $\theta$ variations and mode suppression}
\begin{figure}[t]
	\flushleft
		\includegraphics[width=8.5cm]{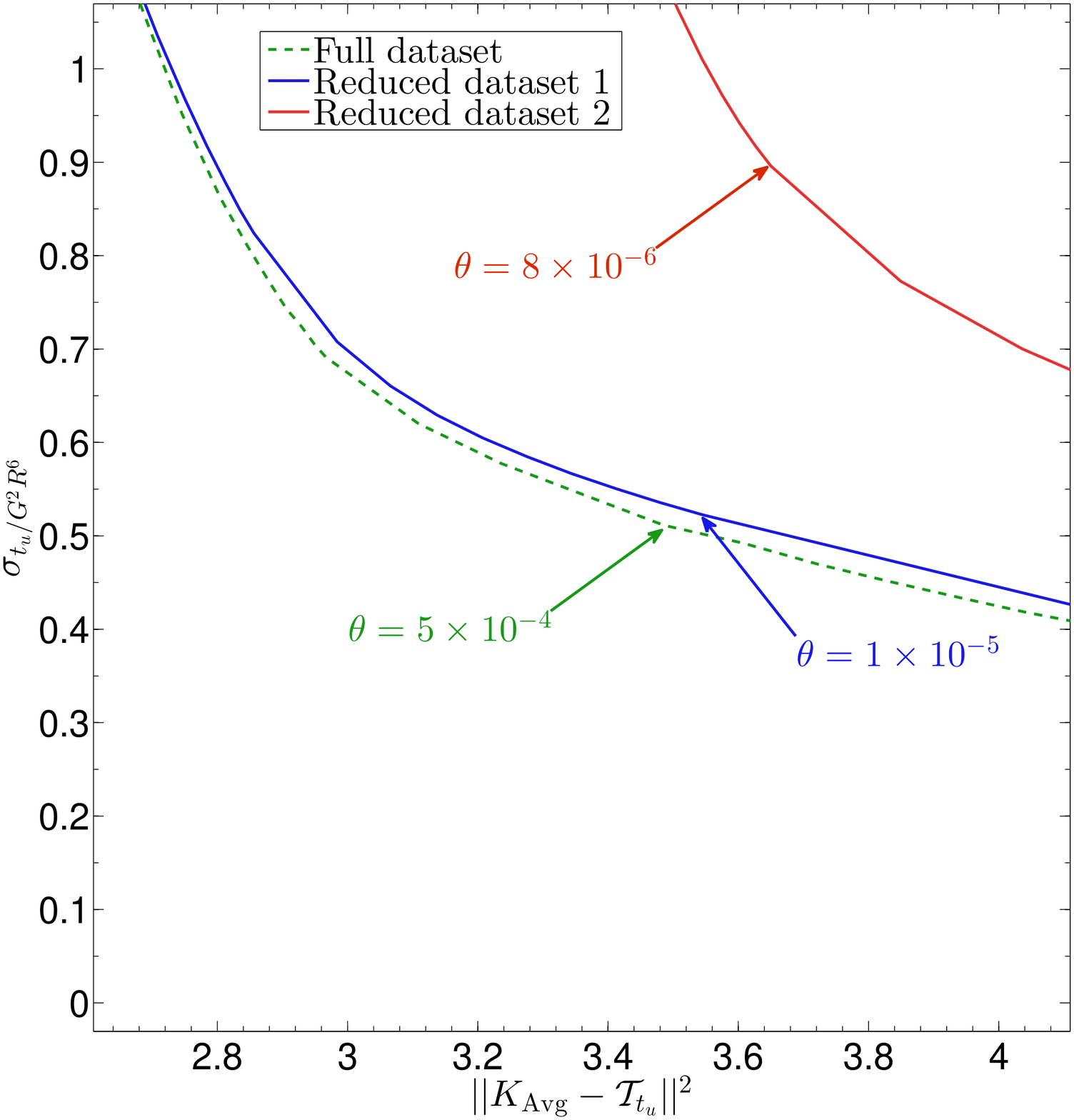}
	\caption{Trade-off curves for the $t_{u}$ inversion using SOLA for the full dataset and for the dataset not using the three octupole modes with lowest radial order.}
		\label{FigTradeoff}
\end{figure}

As we stated in the previous section, the SOLA inversion is a compromise between precision and accuracy. This compromise is materialized by what is called a trade-off curve. It presents the accuracy of the result, in the form of the fit of the target function in abscissa plotted against the observational error amplification in ordinate for different values of the $\theta$ parameter. An example of a trade-off curve for the full-set of observation is plotted in Fig. \ref{FigTradeoff}.

As we can see, this trade-off curve is an L-shaped curve and it is also quickly understood that the optimal value of the free parameter $\theta$ is found nearby the edge of the trade-off curve. This position is associated with the best compromise achievable between precision and accuracy given a set of observational data. For the particular case of $16$CygB, we started with values of $\theta=10^{-5}$ and found out that values around $5 \times 10^{-5}$ were better in terms of compromise between precision and accuracy. This is indeed seen in the plot of the trade-off curve were we zoomed on regions associated with $\theta=10^{-5}$. The green line is the trade-off curve obtained with the full set of data while the blue curve is associated with the trade-off curve when the $\ell=3$ and $n=14,15$ and $16$ modes have been suppressed from the data set. The green vertical line indicates the position on the green trade-off curve (full data set) associated with $\theta=5 \times 10^{-5}$ while the blue vertical line indicates the position on the blue trade-off curve (restricted data set) for $\theta=10^{-5}$. We can see that both positions are very close to each other in terms of error bars and fit of the target function values. However, the fact that the blue curve is always above and to the right of the green curve means that the compromise achieved with the restricted dataset will always be sub-optimal when compared to the compromise achieved with the full dataset. One can also see that the changes in error bars are quite quick when reducing $\theta$ to lower values. For example, if one considers the initial value of $\theta=10^{-5}$, the error bars are $30 \%$ larger then at $\theta=5 \times 10^{-5}$ (which is even more striking then the example given in Fig. \ref{figError}).

\begin{figure}[t]
	\flushleft
		\includegraphics[width=8.5cm]{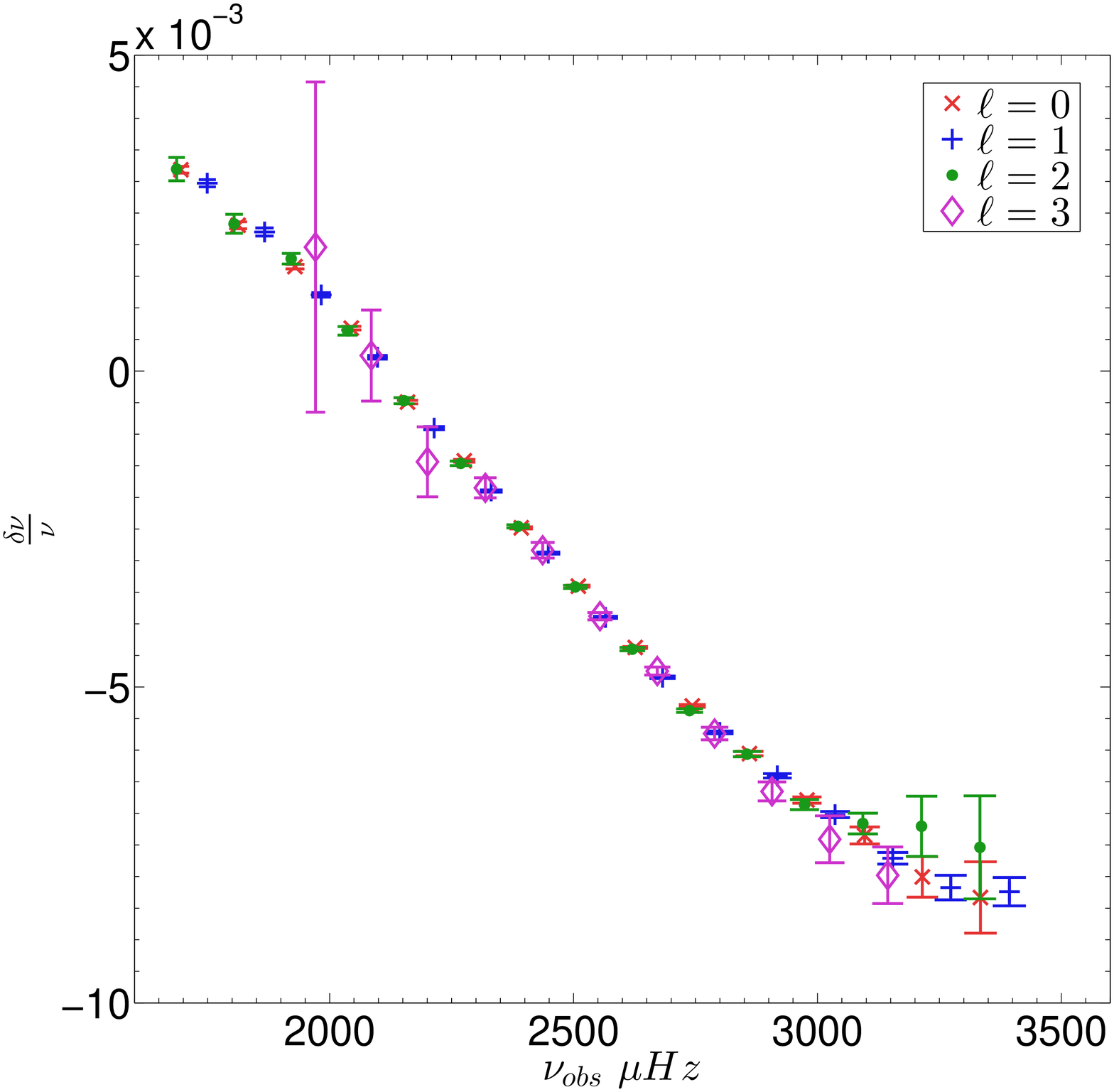}
	\caption{Relative frequency differences plotted with respect to the observed frequencies.}
		\label{FigDiff}
\end{figure}

To illustrate the reason why we tried to eliminate the modes associated with $\ell=3$ and $n=14,15$ and $16$, we plot in Fig. \ref{FigDiff} the individual relative frequency differences with increasing frequencies. It can be seen that these three modes are well fitted and have larger error bars, this is why suppressing them helped us find a better compromise for the inversion technique. However, as stated above, this compromise is still sub-optimal in the strict mathematical sense due to the positions of the trade-off curves with respect to each other. 

\begin{figure}[t]
	\flushleft
		\includegraphics[width=9cm]{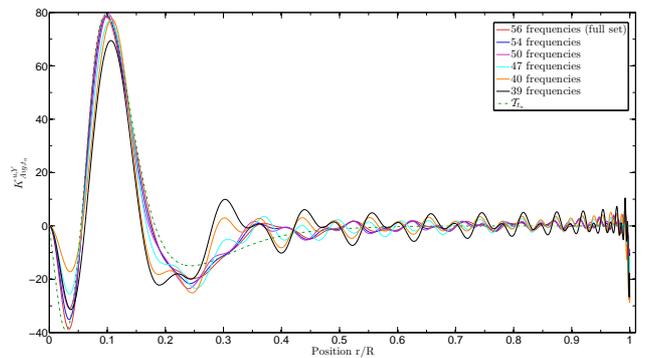}
	\caption{Effects of modes suppression on the averaging kernel of the $t_{u}$ inversion, illustrating the decrease in quality of the target function (in green) fit for various frequency sets. The degradation of the kernel fit is associated with instabilities in the $t_{u}$ values given in Table \ref{tabtu}.}
		\label{figModes}
\end{figure}

Further illustrations are provided in Fig. \ref{figModes} and Table \ref{tabtu}. In Fig. \ref{figModes}, we illustrate the variation of the fit of the averaging kernel for various sets of observed frequencies. Each fit is also associated with a result in table \ref{tabtu}. For the set of $47$ frequencies and $39$ frequencies, we suppressed modes with small error bars and low $n$, that are known to be used by the inversion. It can be seen that the degradation of the kernel is correlated with a reduction in accuracy and some instability of the inversion results. This is basically due to the fact that each time we change the dataset, we are on a different position on a different trade-off curve. Also, this does not mean that change $\theta$ will always be a solution, because at some point the seismic information will simply be insufficient to infer some diagnostic using the $t_{u}$ inversion. 

\begin{table}[t]
\caption{Degradation of inversion results due to modes suppresion: $t_{u}$ results.}
\label{tabtu}
  \centering
\begin{tabular}{r | c }
\hline \hline
\textbf{Number of modes} & \textbf{$t_{u}$ values $(\theta=10^{-5})$} \\ \hline
$56$ (full set)& $2.97 \pm 0.69$  \\ 
$54$ & $2.94 \pm 0.55$  \\ 
$50$ & $2.92 \pm 0.42$  \\ 
$47$ & $3.47 \pm 0.55$  \\
$40$ &$3.01 \pm 0.32$ \\
$39$ & $3.28 \pm 0.30$ \\
\hline
\end{tabular}
\end{table}
 
\end{document}